\documentclass[12pt]{article}%
\usepackage{amsmath}
\usepackage{amssymb}%
\usepackage{amsfonts}%
\input epsf.tex
\newcommand{\be}{\begin{equation}}
\newcommand{\ee}{\end{equation}}
\newcommand{\bea}{\begin{eqnarray}}
\newcommand{\eea}{\end{eqnarray}}

\begin{document}
\bigskip\begin{titlepage}
\begin{flushright}
UUITP-26/04\\
hep-th/0411172
\end{flushright}
\vspace{1cm}
\begin{center}
{\Large\bf Transplanckian energy production\\
and slow roll inflation\\}
\end{center}
\vspace{3mm}
\begin{center}
{\large
Ulf H.\ Danielsson} \\
\vspace{5mm}
Institutionen f\"or Teoretisk Fysik, Box 803, SE-751 08
Uppsala, Sweden
\vspace{5mm}
{\tt
ulf@teorfys.uu.se \\
}
\end{center}
\vspace{5mm}
\begin{center}
{\large \bf Abstract}
\end{center}
\noindent
In this paper we investigate how the energy density due to a non-standard choice of initial vacuum affects the expansion of the
universe during inflation. To do this we introduce source terms in the Friedmann equations making sure that we respect
the  relation between gravity and thermodynamics. We find that the energy production automatically implies a slow rolling cosmological constant. Hence
we also conclude that there is no well defined value for the cosmological constant in the presence of sources. We speculate
that a non-standard vacuum can provide slow roll inflation on its own.
%unstable
\vfill
\begin{flushleft}
November 2004
\end{flushleft}
\end{titlepage}\newpage

%%%% INTRODUCTION

\section{Introduction}

The main advantage of inflation is that it makes our present day universe to a
very large extent insensitive to the precise initial conditions at the Big
Bang.\footnote{For an excellent review with references, see \cite{liddle}.} In
a sense, inflation replaces initial conditions by dynamics and makes a theory
of the early universe possible. To be more precise, inflation provides a
theory for the effective initial conditions to be imposed on the subsequent
evolution of the universe that takes over when inflation ends.

A particular example of this insensitivity to initial conditions, and the
resulting predictive power, is the role quantum fluctuations play in the
theory of inflation. One of the most amazing suggestions put forward during
the past years in cosmology, is that the largest structures of the universe
can be traced back, through the expansion of the universe, to microscopical
quantum fluctuations occuring during the inflationary era. A natural question
to ask, in this context, is how the vacuum for these fluctuations is supposed
to be chosen. For inflation to have any predictive power this choice must be
highly restricted. Luckily, the key feature of inflation, the accelerated
expansion of the universe, provides an answer. If we follow a given
fluctuation backwards in time far enough, its wavelength will eventually
become arbitrarily smaller than the horizon radius. This means that deviations
from Minkowsky space will become less and less important with respect to
defining the vacuum, and the vacuum becomes, in this way, essentially unique.
This vacuum, sometimes called \textit{the Bunch-Davies vacuum, }is what one
should use when finding out what the predictions of inflation are. The fact
that a unique vacuum is picked out is an important property of inflation and
is one of several examples of how inflation does away with the need to choose
initial conditions.

However, the argument relies on an ability to follow a given mode to
infinitely small scales which is not how it\ is expected to work in the real
world. After all, it is generally believed that there exists a fundamental
scale -- Planckian or stringy -- where physics could be completely different
from what we are used to, and where we have very little control of what is
happening. How does this affect the argument that the inflationary vacuum is
unique? Could there be effects of new physics which will affect the
predictions of inflation? In particular, could there be appreciable changes in
the expected nature of the CMBR fluctuations?

Several groups have investigated ways of modifying high energy physics in
order to look for such modifications, see, e.g., [2-25]. One approach is to
modify the dispersion relation at high energy -- we will come back to a
particular toy model of this kind below. Another possible approach, following
\cite{Danielsson:2002kx}, is to by hand impose initial conditions (in
principle coming from unknown high energy physics), corresponding to a
particular inflaton quantum state, and to investigate what the effects, if
any, there are on the CMBR. To proceed along this direction, we need to find
out \textit{when} to impose the initial conditions for a mode with a given
(constant) comoving momentum $k$. To do this, we use conformal time, given by
$\tau=-\frac{1}{aH}$, and note that the physical momentum $p$ and the comoving
momentum $k$ are related through
\begin{equation}
k=ap=-\frac{p}{\tau H}.
\end{equation}
We impose the initial conditions when $p=\Lambda$, where $\Lambda$ is the
energy scale, maybe the Planck scale or possibly the string scale, where
fundamentally new physics become important. The conformal time when the
initial condition is imposed is then given by
\begin{equation}
\tau_{0}=-\frac{\Lambda}{Hk}.\label{eta0}%
\end{equation}
As we see, different modes will be created at different times, with a smaller
linear size of the mode (larger $k$) implying a later time.

The basic idea is that we do not know what happens at higher energies, or
shorter wavelengths, than $\Lambda,$ and that we therefore are forced to
encode our ignorance in terms of initial conditions when the modes enter into
the regime that we understand. The unknown high energy physics is usually
referred to as \textit{transplanckian}, and the hope is that, e.g., string
theory eventually will give us the means to derive these effective initial
conditions. It is well known that the choice of vacuum is a highly non trivial
issue in a time dependent background. Without knowledge of the transplanckian
physics we can only list various possibilities and investigate whether there
is a typical size or signature of the new effects. In \cite{Danielsson:2002kx}
an argument was provided for how to choose a natural vacuum and that this
vacuum in general can be expected to differ from the Bunch Davies vacuum.
Expanding the rescaled inflaton field, $\mu=a\phi$, to lowest adiabatic order
as%
\begin{equation}
\mu_{k}=\frac{\alpha_{k}}{\sqrt{2k}}e^{-ik\eta}+\frac{\beta_{k}}{\sqrt{2k}%
}e^{ik\eta},
\end{equation}
it was argued in \cite{Danielsson:2002kx} that the Bogolubov coefficients have
the natural order
\begin{equation}
\beta\sim\frac{H}{\Lambda}.
\end{equation}
As discussed in \cite{Danielsson:2002qh}, the initial condition approach to
the transplanckian problem allows for a discussion of many of the
transplanckian effects in terms of \textit{the }$\alpha$\textit{-vacua.} These
vacua have been known since a long time, \cite{alpha}, and corresponds to a
family of vacua in de Sitter space which respects all the symmetries of the
space time. With the use of this input it was shown in
\cite{Danielsson:2002kx} how this leads to a simple formula describing a
modulated spectrum for the CMBR. Luckily, the possible changes of the standard
inflationary predictions are small enough not to ruin the framework, but not
so small that they are obviously observationally irrelevant.

One possible concern that one can have with these vacua is in what way the
excess energy density that they represent will affect the evolution of the
universe. This is the main issue we will investigate in the rest of this paper.

\section{The problem of back reaction}

The non-standard vacuum contributes an extra energy density that, potentially,
could back react on the geometry and change the way the universe
expands.\footnote{Following standard, and not obviously justified, procedure
we will assume that the energy of the Bunch Davies vacuum already is
subtracted off and the only contribution we need to be considering is the
energy of the excitations. This is a common starting point for all discussions
on inflation.} Assuming an inflationary cosmology where the inflaton dominates
the energy density, we must make sure that the contribution from the vacuum
energy is negligible \cite{Tanaka:2000jw}. The energy density is given by%
\begin{equation}
\rho_{\Lambda}\sim\int_{0}^{\Lambda}dpp^{3}\left\vert \beta\right\vert
^{2}\sim\Lambda^{2}\left\vert \beta\right\vert ^{2},
\end{equation}
assuming scale independent Bogolubov coefficients, where we have assumed that
there is no contribution to the energy density above the fundamental scale
$\Lambda$. This energy density needs to be compared with the energy density
driving inflation which is given by%
\begin{equation}
\rho\sim M_{pl}^{2}H^{2}.
\end{equation}
If one wants to avoid the excess vacuum energy to affect the expansion rate of
the universe, one must make sure that $\rho_{\Lambda}\ll\rho$. Luckily, this
is the case for the vacua argued to be natural in \cite{Danielsson:2002kx}.
From above it follows that we can ignore the effect of the vacuum energy
provided that $\Lambda\ll M_{pl}$, which has to be assumed in these models for
other reasons. Furthermore, with $\Lambda$ as the string scale, this is a
rather natural requirement. However, as we will see further on, there is more
to the story than this, and we will find that the vacuum energy can play a
quite important role under the right circumstances.

Another objection was put forward in \cite{Starobinsky:2002rp}, where it was
argued that\ the same physics selecting a non standard vacuum during inflation
should still be at work today. The value of $\beta$ would certainly be
expected to be much smaller, presumably determined by the present day Hubble
constant, and the expected energy density would be $\rho\sim\Lambda
^{2}H_{today}^{2}$. Nevertheless, one would expect this energy to be present
in the form of ultra high energy particles which, through their interactions
with other matter, would produce also lower energy gamma and cosmic rays. The
estimated rates are a few order of magnitude higher than what is measured, and
in \cite{Starobinsky:2002rp} this was used to argue for limits on the possible
vacua such that effects on the CMBR are excluded.

In \cite{Brandenberger:2004kx} it was proposed that even a substantial energy
density arising from vacuum fluctuations (not necessarily with the value of
$\beta$ used in this paper) might not destroy inflation but instead act like a
renormalization of the inflationary cosmological constant. This was shown in
the framework of a particular transplanckian model with a modified dispersion
relation. The focus in this model is on a scalar field with a modified
dispersion relation at high energies, see figure 1. 
\begin{figure}
\begin{center}
\centering
\epsfysize=7cm
\leavevmode
\epsfbox{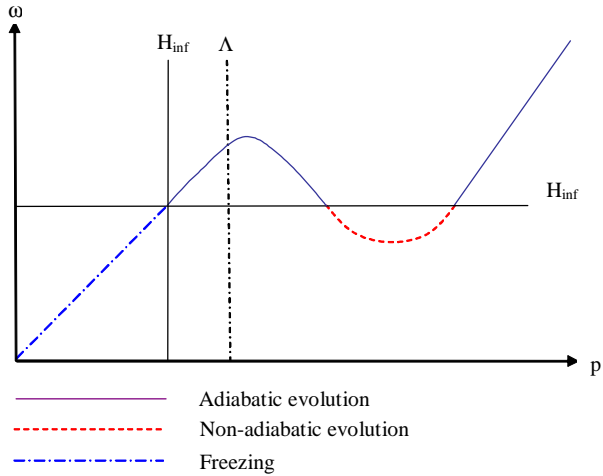}
\end{center}   
\caption[]{\small The evolution (not to scale) of a mode with a modified dispersion
relation during inflation. The effective initial conditions imposed at
$p=\Lambda$ will be the ones of an excited state.}
\end{figure}

At low energies the
dispersion relation is linear with frequency increasing with momentum in the
standard fashion. In an intermediate energy range the frequency decreases with
momentum, and then at high energies it increases again. The main idea is that
the initial state at really high energies is the adiabatic one. As the
universe expands, and the energy redshifts, the frequency of a mode remains,
for high energies, larger than the Hubble constant. This corresponds to an
adiabatic evolution and the vacuum does not change. In the intermediate regime
the frequency drops due to the non standard dispersion relation. As a
consequence it becomes lower than the Hubble constant and the evolution is no
longer adiabatic. When the universe has expanded further, and the mode
redshifted even more, the frequency again becomes larger than the Hubble
constant and another adiabatic phase can begin. At this point, however, the
state of the field no longer coincides with the adiabatic vacuum.%

Furthermore, in this scenario, it is very natural to expect, contrary to the
argument in \cite{Starobinsky:2002rp}, a very different result for the energy
production today. The only thing we need to make sure is that the Hubble scale
today is lower than the minimum of the kink on the dispersion relation. This
case is shown in figure 2. 
\begin{figure}
\begin{center}
\centering
\epsfysize=7cm
\leavevmode
\epsfbox{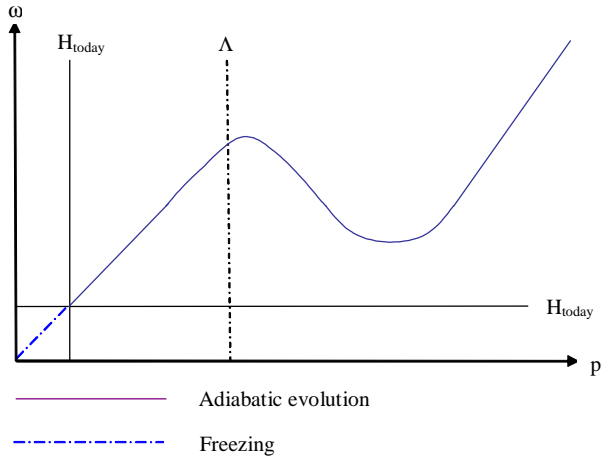}
\end{center}   
\caption[]{\small The evolution (not to scale) of a mode with a modified dispersion
relation today. The effective initial conditions when $p=\Lambda$ will be the
ones of the adiabatic vacuum.}
\end{figure}
As a consequence, there is no non-adiabatic
evolution at transplanckian energies and the vacuum remains the adiabatic one.
Hence there is no excessive production of high energy radiation at the present times.%
In the framework of \cite{Danielsson:2002kx} the new vacuum is used as an
initial state once the mode has redshifted out of the regime with a non
standard dispersion relation, as is depicted in figure 1. In figure 2 we have
a situation where the standard vacuum is picked out.

In the rest of the paper we will show how the same general conclusions about
the influence of the extra vacuum energy during inflation, can be reach in the
framework of \cite{Danielsson:2002kx} where we do not make any explicit
assumptions about the transplanckian physics. It is reassuring though, that a
precise example, as the one of \cite{Brandenberger:2004kx}, exists showing the
full consistency of the argument.

\bigskip

\section{Sourced Friedmann equations}

\bigskip

In order to address the problem of back reaction we will have to introduce the
energy density of the non-standard vacuum into the Friedmann equations, and
for consistency, we will also need to include a source term due to the
continuos creation of modes.\footnote{The necessity of source terms has also
been discussed in \cite{Keski-Vakkuri:2003vj}.} From the point of view of the
specific model considered in \cite{Brandenberger:2004kx} the source would
correspond to the non-adiabatic phase where excited modes are created.
Contrary to \cite{Brandenberger:2004kx}, we will, in our simplified framework,
be able to solve for the back reaction in some detail. In fact, as will become
clear further on, we will see how the vacuum energy can drive inflation all on
its own.

In order to generalize the Friedmann equations to include sources, we will
find it useful to discuss the equations from the thermodynamical point of view
taken in \cite{Jacobson:1995ab}.

\subsection{A thermodynamical approach to the Friedmann equations}

\bigskip

Given the connection between black hole physics and thermodynamics first
revealed by Bekenstein in the seventies,\ \cite{Bekenstein:ax}, it is tempting
to speculate about a deeper connection between thermodynamics and gravity in
general. Along this line of thought, it was argued in \cite{Jacobson:1995ab}
that the gravitational Einstein equations can be derived through a
thermodynamical argument using the relation between area and entropy as input.
In the simplified cosmological setting relevant for our analysis, the
corresponding argument was given in \cite{Frolov:2002va}.

The starting point is the relation between the area of the horizon and entropy
for a black hole given by%
\begin{equation}
S=\frac{M_{p}^{2}}{4}A.
\end{equation}
In an expanding universe, at least in the case of accelerated expansion, the
cosmological horizon determined by the Hubble constant plays a similar role as
the horizon of a black hole, \cite{GibbHawk:1977}. We therefore assign an
entropy to the horizon according to%
\begin{equation}
S=\frac{\pi M_{p}^{2}}{H^{2}}.
\end{equation}
We now proceed with deriving the Friedmann equations describing the time
evolution of the universe with the entropy relation as a starting point. To do
this, we use the standard relation between flow of heat and entropy, $dQ=TdS$.
The flow of heat out through the horizon is then related to a flow of entropy
given by
\begin{equation}
\dot{Q}=\dot{S}T=A\left(  \rho+p\right)  ,
\end{equation}
where
\begin{equation}
T=\frac{H}{2\pi}.
\end{equation}
Using that the entropy can be expressed in terms of the horizon area and the
Hubble constant, we find%
\begin{equation}
\dot{H}=-\frac{4\pi}{M_{p}^{2}}\left(  \rho+p\right)  ,\label{eq:h0prick}%
\end{equation}
which, indeed, is one of the Friedmann equations.

To completely specify the time evolution we also need the continuity equation%
\begin{equation}
\overset{\cdot}{\rho}+3H\left(  \rho+p\right)  =0,
\end{equation}
which, combined with (\ref{eq:h0prick}), give another of the Friedmann
equations, i.e.,%

\begin{equation}
H^{2}=\frac{8\pi}{3M_{p}^{2}}\rho.\label{eq:h02}%
\end{equation}
Usually the two Friedmann equations together with the continuum equation are
viewed on an equal footing keeping in mind that only two of them are
independent. However, from our thermodynamical point of view, there is an
important difference between the various choices. (\ref{eq:h02}) is obtained
from (\ref{eq:h0prick}) using integration and there is, therefore, a
corresponding constant of integration, \textit{the cosmological constant},
which does not appear in the basic equations (\ref{eq:h0prick}). The usual
interpretation is that the cosmological constant corresponds to matter with
$p=-\rho$. However, from our thermodynamical point of view, it is more natural
to view the cosmological constant as part of the initial conditions.

While the above argument was carried out for the specific example of a
FRW-cosmology, it was given in all generality in \cite{Jacobson:1995ab}, where
it was shown that the thermodynamical approach to gravity required%
\begin{equation}
8\pi GT_{\mu\nu}=R_{\mu\nu}+fg_{\mu\nu},
\end{equation}
with $f$ an arbitrary function. If one then demands that the energy momentum
tensor is conserved, one finds that $f=-\frac{R}{2}+\Lambda_{c}$ with
$\Lambda_{c}$ as the cosmological constant, and the Einstein equations follows.

The thermodynamical approach to gravity might be considered as a curious
observation, and nothing more. In the following we will see, however, that
this point of view makes it a bit easier to think about an expanding universe
in the presence of sources.

\bigskip

\subsection{With sources}

\bigskip

How does the above derivation of the Friedmann equations change if $T_{\mu\nu
}$ is \textit{not} conserved? From a thermodynamical point of view it is
obvious that we should make sure that we keep (\ref{eq:h0prick}) unchanged.
This is the equation that relates the change in area of the horizon with the
local flow of matter and energy, and is required by the thermodynamical
interpretation of horizon area. On the other hand, we have, in the presence of
a source $q$,%
\begin{equation}
\overset{\cdot}{\rho}+3H\left(  \rho+p\right)  =q,
\end{equation}
which is easily seen to lead to%
\begin{equation}
H^{2}=\frac{8\pi}{3M_{p}^{2}}\rho-\frac{8\pi}{3M_{p}^{2}}\int^{t}qdt.
\end{equation}
It is clear that (\ref{eq:h0prick}) should remain the same even in the
presence of sources, but we see that this is not at all true for (\ref{eq:h02}).

While we will not need this in the present paper, it is trivial to perform a
general analysis for other backgrounds. We write the generalized continuity
equation and Einstein equation as%
\begin{align}
8\pi G\nabla^{\mu}T_{\mu\nu}  & =g_{\mu\nu}\nabla^{\mu}h\\
8\pi GT_{\mu\nu}  & =G_{\mu\nu}+hg_{\mu\nu}.
\end{align}
With the energy momentum tensor on the form%
\begin{equation}
T_{\mu\nu}=\left(  \rho+p\right)  V_{\mu}V_{\nu}-pg_{\mu\nu},
\end{equation}
it is easy to see that we can define a new energy momentum tensor that
\textit{is} conserved, with new energy and pressure given by%
\begin{align}
\tilde{\rho}  & =\rho+h\\
\tilde{p}  & =p-h
\end{align}
In our cosmological example we have%
\begin{equation}
\tilde{\rho}=\rho-\int^{t}qdt
\end{equation}
After this general discussion, let us turn to the transplanckian problem.

\section{The effect of transplanckian energy production}

\bigskip

\subsection{General discussion}

Let us now make use of the results of the previous section. In addition to the
non-standard vacuum, we allow for the existence of ordinary matter. The two
contributions obey the continuity equations%

\begin{align}
\overset{\cdot}{\rho}_{\Lambda}+3H\left(  \rho_{\Lambda}+p_{\Lambda}\right)
& =q\\
\overset{\cdot}{\rho}_{m}+3H\left(  \rho_{m}+p_{m}\right)   & =0,
\end{align}
where we have allowed for energy production in the vacuum sector, and we have
assumed that all other matter obey the standard equations.\footnote{In
\cite{Horvat:2004vn} source terms were introduced in a different way in order
to sustain an energy density $H^{2}\Lambda^{2}$ motivated from holography. In
that case the source terms represented energy transfer between different
components of matter with the total energy momentum tensor conserved. In our
case we have a net creation of energy, possibly (but not necessarily) caused
by transplanckian non-adiabaticity.} We assume the equations of state to be
given by%
\begin{align}
p_{\Lambda}  & =w_{\Lambda}\rho_{\Lambda}\\
p_{m}  & =w_{m}\rho_{m},
\end{align}
We then impose the Friedmann equation%
\begin{equation}
\dot{H}=-\frac{4\pi}{M_{p}^{2}}\left(  \rho_{\Lambda}+p_{\Lambda}+\rho
_{m}+p_{m}\right)  ,\label{eq:hprick}%
\end{equation}
where we will put $w_{\Lambda}=\frac{1}{3}$, and keep $w_{m}$ arbitrary.
Integrating the equations gives%
\begin{equation}
H^{2}=\frac{8\pi}{3M_{p}^{2}}\left(  \rho_{\Lambda}+\rho_{m}\right)
-\frac{8\pi}{3M_{p}^{2}}\int^{t}qdt,\label{eq:h2q}%
\end{equation}
as discussed in the previous section.

Let us now consider the specific case of vacuum fluctuations with the
characteristic values of the Bogolubov mixing given by%
\begin{equation}
\left\vert \beta_{k}\right\vert ^{2}\sim\frac{H^{2}}{\Lambda^{2}}.
\end{equation}
Assuming an essentially constant $H$, and integrating over all energies up to
the cutoff scale yields%
\begin{equation}
\rho_{\Lambda}\sim\int_{0}^{\Lambda}dpp^{3}\frac{H^{2}}{\Lambda^{2}}%
\sim\Lambda^{2}H^{2}.
\end{equation}
For convenience we redefine $\Lambda$ such that
\begin{equation}
\rho_{\Lambda}=\frac{3\Lambda^{2}H^{2}}{8\pi}.\label{eq:rholambda}%
\end{equation}
Actually, we will have to be a bit more careful than this. Since $H$ will be
changing with time, i.e. decrease, we must take this into account when
calculating the vacuum energy density. Modes with low momenta were created at
earlier times when the value of $H$ were larger, and there will be a slight
enhancement in the way these modes contribute to the energy density. We
therefore write%
\begin{equation}
\rho_{\Lambda}\left(  a\right)  =\frac{3}{2\pi}\int_{\varepsilon}^{\Lambda
}dpp^{3}\frac{H^{2}\left(  \frac{ap}{\Lambda}\right)  }{\Lambda^{2}}=\frac
{3}{2\pi}\frac{1}{\Lambda^{2}a^{4}}\int_{a_{i}}^{a}dxx^{3}H^{2}\left(
x\right)  ,
\end{equation}
where we have introduced a low energy cutoff corresponding to the present
energy of modes that started out at $\Lambda$ at a time when the Hubble
constant was as small as $H_{i}$.\footnote{The exact expressions for the modes
are modified at low momenta in an expanding universe. This yields corrections
typically suppressed by further orders of $\frac{H^{2}}{\Lambda^{2}}$.}

If we take a derivative of the energy density with respect to the scale factor
and use $\frac{d}{da}=\frac{1}{aH}\frac{d}{dt}$, we find%
\begin{equation}
\dot{\rho}_{\Lambda}+4H\rho_{\Lambda}=\frac{3}{2\pi}\Lambda^{2}H^{3}%
,\label{eq:contQ}%
\end{equation}
and we can conclude that the source term is given by%
\begin{equation}
q=\frac{3}{2\pi}\Lambda^{2}H^{3}.
\end{equation}

To proceed it is convenient to write (\ref{eq:contQ}) in the form%
\begin{equation}
\frac{d}{da}\left(  a^{4}\rho_{\Lambda}\right)  =\frac{3}{2\pi}\Lambda
^{2}a^{3}H^{2},
\end{equation}
and take a derivative of (\ref{eq:hprick}) with respect to the scale factor to
obtain%
\begin{equation}
a^{2}HH^{\prime\prime}+a^{2}H^{\prime2}+5aHH^{\prime}=-\frac{8\Lambda^{2}%
}{M_{p}^{2}}H^{2}-\frac{4\pi}{M_{p}^{2}}\left(  1+w_{m}\right)  \left(
1-3w_{m}\right)  \rho_{m},
\end{equation}
where we have used that $\rho_{m}\sim a^{-3\left(  1+w_{m}\right)  }$. The
general solution is easily found to be%
\begin{equation}
H^{2}=C_{1}^{2}a^{-2n_{1}}+C_{2}^{2}a^{-2n_{2}}+\frac{8\pi}{3M_{p}^{2}}%
\frac{\left(  1+w_{m}\right)  \left(  1-3w_{m}\right)  }{\left(
1+w_{m}\right)  \left(  1-3w_{m}\right)  -\frac{16\Lambda^{2}}{3M_{p}^{2}}%
}\rho_{m},\label{eq:h2c1c2}%
\end{equation}
where $C_{1,2}$ are constants of integration, and%
\begin{equation}
n_{1,2}=1\pm\sqrt{1-\frac{4\Lambda^{2}}{M_{p}^{2}}}.
\end{equation}
In the limit $\Lambda\rightarrow0$ (implying that the vacuum energy is
removed) we find%
\begin{equation}
H^{2}=C_{1}^{2}+C_{2}^{2}a^{-4}+\frac{8\pi}{3M_{p}^{2}}\rho_{m}.
\end{equation}
That is, $C_{1}$ gives a cosmological constant, while $C_{2}$ corresponds to
radiation, both of which can be absorbed into $\rho_{m}$. More interesting, is
the case when $\Lambda\neq0$. We see, then, that neither a cosmological
constant nor, which is less expected, a radiation component survives in
(\ref{eq:h2c1c2}) since $\left(  1+w_{m}\right)  \left(  1-3w_{m}\right)  =0$
for $w_{m}=-1$ or $w_{m}=1/3$. Instead, they both appear through constants of
integration. Furthermore, due to the source term, the way the two components
depend on the scale factor is changed. For small $\frac{\Lambda^{2}}{M_{p}%
^{2}}$, we find%
\begin{equation}
\left\{
\begin{array}
[c]{l}%
n_{1}\sim2-\frac{2\Lambda^{2}}{M_{p}^{2}}\\
n_{2}\sim\frac{2\Lambda^{2}}{M_{p}^{2}}.
\end{array}
\right.
\end{equation}
The first corresponds to a radiation component which is decaying with redshift
a bit slower than usual, while the other is more like a cosmological constant
that is slowly decreasing. 

Let us finally consider the case with only radiation, that is $w_{m}=1/3$, in
some more detail. We then have%
\begin{equation}
H^{2}=C_{1}^{2}a^{-2n_{1}}+C_{2}^{2}a^{-2n_{2}}\label{eq:h2rad}%
\end{equation}
In particular, let us consider the initial moment when $H=H_{i}$. At that time
we have, by definition, $\rho_{\Lambda}=0$. Note that $\rho_{\Lambda}$ always
correspond to energy created after the time of an, arbitrary, initial scale
factor $a_{i}$. But how much radiation is already present? This is given
directly by (\ref{eq:hprick}), since the flow of radiation out through the
horizon dictates the way the Hubble constant changes. We find%
\begin{equation}
\frac{a}{2}\frac{d}{da}H^{2}=-\frac{16\pi}{3M_{p}^{2}}\rho_{m},
\end{equation}
from which we conclude%
\begin{equation}
\frac{8\pi}{3M_{p}^{2}}\rho_{m}=\frac{n_{1}}{2}C_{1}^{2}a_{i}^{-2n_{1}}%
+\frac{n_{2}}{2}C_{2}^{2}a_{i}^{-2n_{2}}.\label{eq:rhorad}%
\end{equation}
That is, for any $a_{i}$, we can read off the amount of radiation present at
that scale factor. It is not very surprising that the evolution with the scale
factor does not go as $1/a^{4}$, since, after all, we have made sure that
there is a continuos creation of matter. It is interesting to consider the
difference between (\ref{eq:h2rad}) and (\ref{eq:rhorad}). This is given by%
\begin{equation}
\frac{8\pi}{3M_{p}^{2}}\rho_{\Lambda_{cc}}=C_{1}^{2}\left(  1-\frac{n_{1}}%
{2}\right)  a^{-2n_{1}}+C_{2}^{2}\left(  1-\frac{n_{2}}{2}\right)  a^{-2n_{2}%
},
\end{equation}
and would be expected to be identified with a cosmological constant. This is,
however, true only in the limits where $\Lambda=0$.
Otherwise we obtain a cosmological constant that is slowly decaying. It is
important to observe that \textit{in the presence of sources one can not
assign an unambiguous value to the cosmological constant. }This is one of the
main conclusions of the paper. We find that a fixed dimensionful cosmological
constant, is effectively replaced by a dimensionless parameter determining the
running, given by the ratio of a fundamental scale and the Planck
scale.\footnote{One should not that a similar relaxation of the cosmological
constant has been observed in \cite{Mottola:1984:ar}. In that case, however,
it was argued to be due to an energy density present already in the Bunch
Davies vacuum. The timescale was given by the mass of the fluctuating field.}

\section{Conclusions and speculations}

\bigskip

In this paper we have investigated in what way a non-standard vacuum affects
the expansion of the universe during inflation due to the extra energy
density. To do this we have found it necessary to include a source term in the
continuity equation that also affected one of the Friedmann equations. A nice
framework for understanding how to make the appropriate modifications is
provided by the work of \cite{Jacobson:1995ab}.

We have found, in the presence of sources (e.g. due to non-adiabatic
transplanckian physics), that a cosmological constant is replaced by a running
quantity. These results are consistent with, and generalize, what was found in
\cite{Brandenberger:2004kx}. As a consequence, a non-standard vacuum can yield
an inflationary slowroll all on its own even with out a nontrivial
inflationary potential. The results suggests that in searching for
phenomenologically viable inflationary models, the choice of initial vacuum
(or equivalently, the details of transplanckian physics) could play a similar
role as the shape of the inflaton potential. It would be interesting to find
models with more detailed phenomenology and also investigate the relation with
detectable modulations of the CMBR. This is left for future work.

\bigskip

\section*{Acknowledgments}

The author is a Royal Swedish Academy of Sciences Research Fellow supported by
a grant from the Knut and Alice Wallenberg Foundation. The work was also
supported by the Swedish Research Council (VR).

\end{document}